\documentclass[tikz]{article}
\usepackage{spconf,amsmath,graphicx}
\usepackage{booktabs}
\newcommand{\argmin}{\mathop{\mathrm{arg\,min}}}

\usepackage{enumitem}
\setlist{nosep, leftmargin=14pt}

\usepackage{mwe} %

\usepackage{amssymb}
\usepackage[dvipsnames]{xcolor}
\usepackage{multirow}
\usepackage{pgfplots}
\usepackage{subcaption}
\usepackage{url}
\usepackage{soul}

\newcommand{\std}[1]{\scriptsize{$\pm$#1}}

\def\L{{\cal L}}

\title{Contrastive learning for regression in multi-site brain age prediction}
\name{Carlo Alberto Barbano$^{\star\dagger}$\thanks{Correspondence: carlo.barbano@unito.it} \quad Benoit Dufumier$^{\star\ddagger}$ \quad Edouard Duchesnay$^{\ddagger}$ \quad Marco Grangetto$^{\dagger}$ \quad Pietro Gori$^{\star}$}
\address{$^{\star}$ LTCI, Télécom Paris, IP Paris \quad $^{\dagger}$ University of Turin \quad $^{\ddagger}$ NeuroSpin, CEA, Université Paris-Saclay}
\begin{document}
\maketitle
\begin{abstract}
Building accurate Deep Learning (DL) models for brain age prediction is a very relevant topic in neuroimaging, as it could help better understand  neurodegenerative disorders and find new biomarkers. To estimate accurate and generalizable models, large datasets have been collected, which are often multi-site and multi-scanner. This large heterogeneity negatively affects the generalization performance of DL models since they are prone to overfit site-related noise. Recently, contrastive learning approaches have been shown to be more robust against noise in data or labels. For this reason, we propose  a novel contrastive learning regression loss for robust brain age prediction using MRI scans. 
Our method achieves state-of-the-art performance on the OpenBHB challenge, yielding the best generalization capability and robustness to site-related noise.
\end{abstract}


\section{Introduction}
\label{sec:intro}

Brain aging involves complex biological processes, such as cortical thinning, that are highly heterogeneous across individuals, suggesting that people do not age in the same manner. Accurately modeling brain aging at the subject-level is a long-standing goal in neuroscience as it could enhance our understanding of age-related diseases such as neurodegenerative disorders. To this end, brain-age predictors linking neuroanatomy to chronological age have been proposed using Deep Learning (DL)~\cite{peng2021}. 
In order to build accurate biomarker of aging, DL models need large-scale neuroimaging dataset for training, which often involves multi-site studies, partly because of the high cost per patient in each study. 
\begin{figure}
    \centering
    \begin{subfigure}[b]{0.45\linewidth}
        \includegraphics[width=\linewidth]{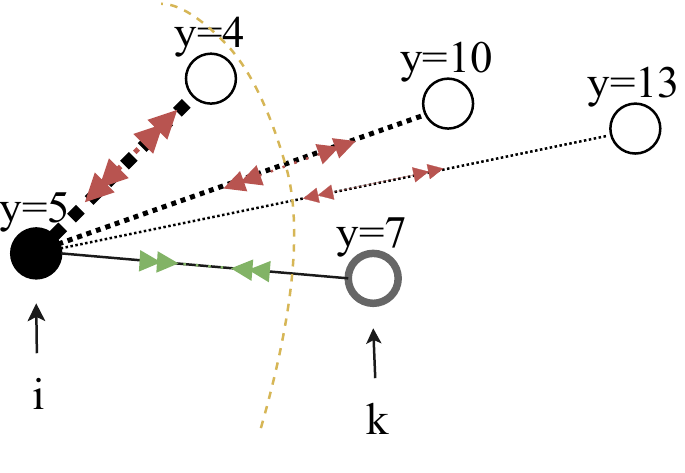}
        \caption{y-aware}
        \label{fig:viz-yaware}
    \end{subfigure}
    \begin{subfigure}[b]{0.45\linewidth}
        \includegraphics[width=\linewidth]{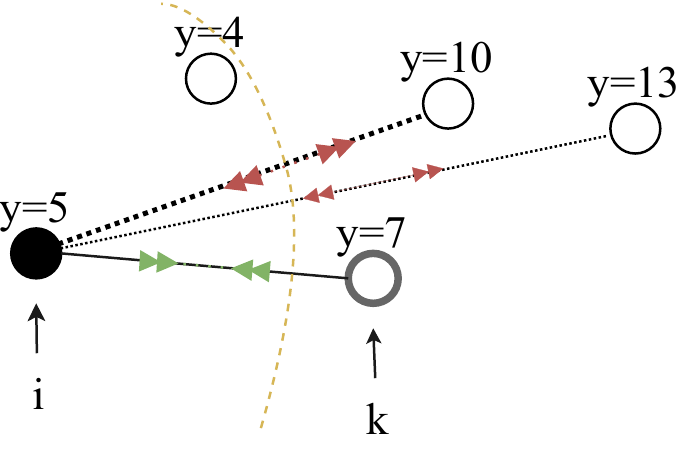}
        \caption{threshold}
        \label{fig:viz-threshold}
    \end{subfigure}
    \begin{subfigure}[b]{0.45\linewidth}
        \includegraphics[width=\linewidth]{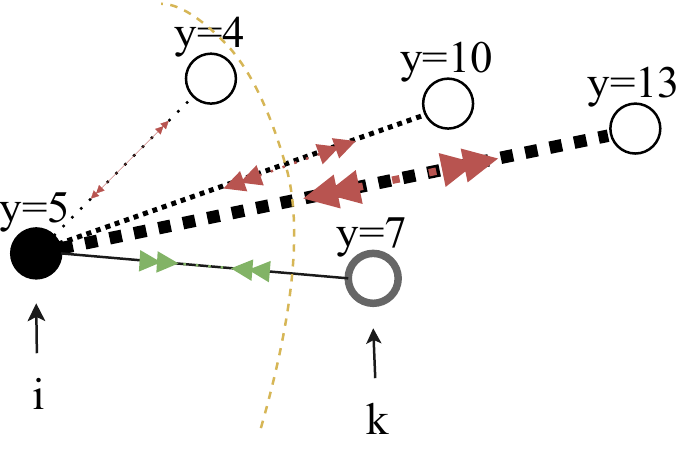}
        \caption{exp}
        \label{fig:viz-expw}
    \end{subfigure}
    \caption{Comparison between different contrastive learning regression losses and their effect on the representations. Samples are aligned ($\gg$ $\ll$) and repelled ($\ll \gg$) with varying strength (line thickness) based on the continuous label $y$. 
    }
    \label{fig:viz-losses}
\end{figure}
Recent works have shown that DL models, and in particular Deep Neural Networks (DNN), largely over-fit site-related noise when trained on such multi-site datasets, notably due to the difference in acquisition protocols, scanner constructors, physical properties such as permanent magnetic field~\cite{wachinger2021detect, glocker2019machine}. This also implies poor generalization performance on data from new incoming sites, highly limiting the applicability of these models to real-life scenarios. 
In order to build more robust and accurate brain age models insensitive to site, the OpenBHB challenge~\cite{dufumier2022openbhb} has been recently released. 
While most DNN used to derive brain age gap are usually trained as standard regressors with the optimization of mean absolute error~\cite{cole2017predicting, Jonsson2019}, Ridge or cross-entropy loss~\cite{peng2021} (if age is binarized), these frameworks do not pay particular care about site-related information during training to produce robust representations of brain imaging data. 
On the other hand, contrastive learning paradigms for DNN training have been recently proposed in various contexts such as supervised~\cite{khosla2020supervised}, weakly-supervised~\cite{tsai2022conditional, dufumier2022rethinking} and unsupervised representation learning ~\cite{chen2020simCLR,barbano2023unbiased}.
More importantly, contrastive learning has been shown to be more robust than traditional end-to-end approaches, such as cross-entropy, against noise in the data or the labels, resulting in better generalizing models~\cite{khosla2020supervised,Graf2021}.
For this reason, in this work, we propose a novel contrastive learning loss for regression in the context of the OpenBHB challenge, where chronological age must be learned without being affected by site-related noise. With our method, we obtain the best results in the official leaderboard. 
Our contributions are twofold:
\begin{itemize}
    \item We propose a novel contrastive learning regression loss for brain age prediction;
    \item We achieve state-of-the-art performance in brain age prediction on the OpenBHB challenge.
\end{itemize}

\section{Method}
\label{sec:method}

Supervised contrastive learning (i.e., SupCon~\cite{khosla2020supervised}) leverages discrete labels (i.e., categories) to define positive and negative samples. Starting from a sample $x_i$, called the \emph{anchor}, and its latent representation $z_i = f(x_i)$, SupCon aligns the representations of all positive samples (i.e. sharing the same class as $x_i$) to $z_i$, while repelling the representations of the negative ones (i.e., different class). The SupCon loss is thus not adapted for regression (i.e., continuous labels), as it is not possible to determine a hard boundary between positive and negative samples. 
All samples are somehow positive and negative at the same time. Given the continuous label $y_i$ for the anchor and $y_k$ for a sample $k$, one could threshold the distance $d$ between $y_i$ and $y_k$ at a certain value $\tau$ in order to create positive and negative samples (i.e., k is positive if $d(y_i, y_k)<\tau$). The problem would then be how to choose $\tau$. Differently, we propose to define a degree of ``positiveness'' between samples using a kernel function $w_k = K(y_i - y_k)$, where $0 \leq w_k \leq 1$. 
Our goal is thus to learn a parametric function $f: \mathcal{X} \rightarrow \mathbb{S}^d$ that maps samples with a high degree of positiveness ($w_k \sim 1$) close in the latent space and samples with a low degree ($w_k \sim 0$) far away from each other. 

\subsection{Contrastive Learning Regression Loss}
\label{sec:method-weakly-contrastive}
In contrastive learning, we look for a parametric mapping function $f$ such that the following condition is always satisfied: $s^-_t - s^+_k \leq 0\quad \forall t,k $, where we denote as $s_k =$ sim($ f(x_i), f(x_k))$ the similarity between the representations of the $k$-th sample and the anchor $i$ (e.g., cosine similarity) and $x^-_t$ and $x^+_k$ are the negative and positive samples respectively. The notion of ``negative'' (dissimilar from the anchor) and ``positive'' (similar to the anchor) samples is thus rooted in the contrastive learning framework. To adapt such a framework to continuous labels, we propose to use a kernel function $w_k$, and we develop multiple formulations, illustrated in Fig.~\ref{fig:viz-losses}. To derive our proposed loss, we employ a metric learning approach, as in ~\cite{barbano2023unbiased}, which allows us to easily add conditioning and regularisation. \\
A first approach would be to consider as ``positive'' only the samples that have a degree of positiveness greater than 0, and align them with a strength proportional to the degree, namely:
\begin{equation}
        \frac{w_k}{\sum_j w_j}(s_t - s_k) \leq 0 \quad  %
        \forall j,k,t \neq k \in A(i) 
\label{eq:yaware-condition}
\end{equation}
where we have normalized the kernel so that the sum over all samples is equal to 1 and we denote with $A(i)$ the indices of samples in the minibatch distinct from $x_i$. Eq.\ref{eq:yaware-condition} can be transformed in an optimization problem using, as it is common in contrastive learning, the $\max$ operator and its smooth approximation \textit{LogSumExp}:
\begin{equation}
    \begin{gathered}
        \argmin_f \sum_k \max(0, \frac{w_k}{\sum_t w_t}\{ s_t - s_k \}_{\substack{t=1,...,N \\ t \neq k}}) =\\ 
        \argmin_f \sum_k \frac{w_k}{\sum_t w_t} \max(0, \{ s_t - s_k \}_{\substack{t=1,...,N \\ t \neq k}}) \\ 
        \approx  \mathcal{L}^{y-aware}= - \sum_k \frac{w_k}{\sum_t w_t} \log \left( \frac{\exp(s_k)}{\sum_{t=1}^N \exp(s_t)}  \right)
    \end{gathered}
\end{equation}
Interestingly, this is exactly the \emph{y-aware} loss proposed in \cite{dufumier2021contrastive} for classification with weak continuous attributes. 
Due to the non-hard boundary between positive and negative samples, both $s_t$ and $s_k$ are defined over the entire minibatch. The kernel $w_k$ is used to avoid aligning samples not similar to the anchor (i.e. $w_k\approx 0$). 
It can be noted that, while the numerator aligns $x_k$, in the denominator, the uniformity term (as defined in \cite{wang_understanding_2020}) focuses more on the closest samples in the representation space: this could be undesirable, as these samples might have a greater degree of positiveness  than the considered $x_k$ (Fig.~\ref{fig:viz-yaware}). %

\noindent To avoid that, we formulate a first extension ($\mathcal{L}^{thr}$) of~\eqref{eq:yaware-condition}, which limits the uniformity term (i.e., denominator) to the samples that are at least more distant from the anchor than the considered $x_k$ in the kernel space (omitting the normalization in the starting condition):
\begin{equation}
\begin{gathered}
    w_k(s_t - s_k) \leq 0 \quad \text{if } w_t - w_k \leq 0 \quad \forall k,t \neq k \in A(i) \\
\mathcal{L}^{thr} = -\sum_k \frac{w_k}{\sum_t \delta_{w_t < w_k} w_t} \log \left( \frac{\exp(s_k)}{\sum_{t\neq k} \delta_{w_t < w_k} \exp(s_t)}\right)
\end{gathered}
\end{equation}
Ideally, $\mathcal{L}^{thr}$ avoids repelling samples more similar than $x_k$. However, it still focuses more on the closest sample ``less positive'' than $x_k$, i.e. $x_t$ s.t $w_t > w_x$ and $w_t \leq w_j\,\, \forall j\neq k$ (Fig.~\ref{fig:viz-threshold}). As noted in~\cite{barbano2023unbiased,khosla2020supervised}, increasing the margin with respect to the closest ``negative'' sample works well for classification, however, it might not be best suited for regression. \\
For this reason, we propose a second formulation ($\mathcal{L}^{exp}$) that takes an opposite approach. 
Instead of focusing on repelling the closest ``less positive'' sample, we increase the repulsion strength for samples proportionally to their distance from the anchor in the kernel space:
\begin{equation}
\begin{gathered}
    w_k[s_t(1-w_t) - s_k] \leq 0 \quad \forall k,t \neq k\in A(i) \\
    \mathcal{L}^{exp} = -\frac{1}{\sum_t w_t}\sum_{k}w_k \log \frac{\exp(s_k)}{\sum_{t \neq k} \exp(s_t(1-w_t))}
\end{gathered}
\end{equation}
In the resulting $\L^{exp}$ formulation, the weighting factor $1 - w_t$ acts like a temperature value, by giving more weight to the samples which are farther away from the anchor in the kernel space (Fig.~\ref{fig:viz-expw}). Also, for a proper kernel choice, samples closer than $x_k$ will be repelled with very low strength ($\sim$0). We argue that this approach is more suited for continuous attributes (i.e., regression task), as it enforces that samples close in the kernel space will be close in the representation space.

\begin{figure*}
    \centering
    \includegraphics[width=\textwidth]{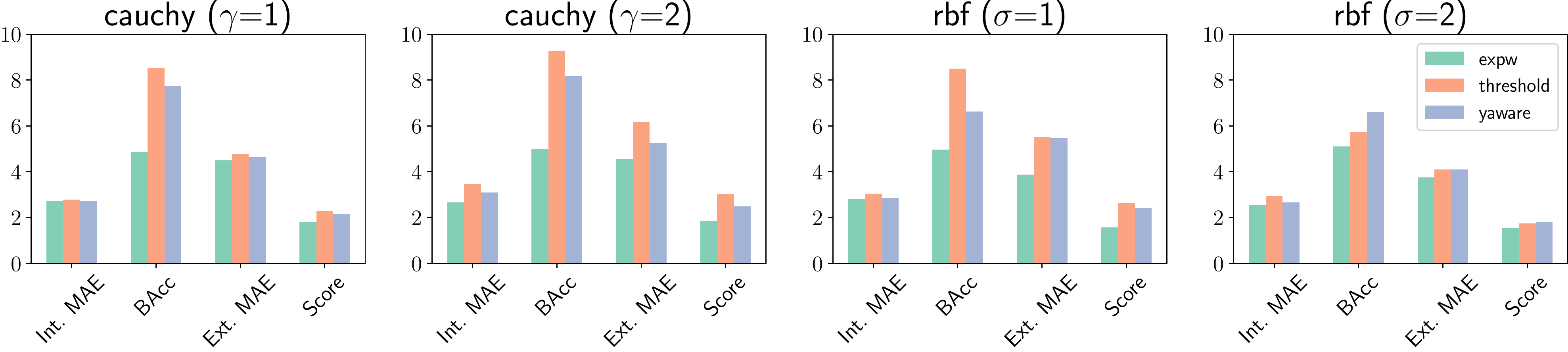}
    \caption{Ablation study of the kernel functions. The Gaussian kernel (rbf) with $\sigma=2$ yields the best generalization results (Ext. MAE) and final score across all three loss functions. We also notice an overall slight improvement in the site balanced accuracy.}
    \label{fig:kernel-ablation}
\end{figure*}

\section{Experimental Data}
We conduct our experiments on the OpenBHB dataset, which contains 5330 3D brain MRI scans from 71 different acquisition sites. Every scan comes from a different subject. The subjects have European-American, European and Asian origins, to promote diversity. Three modalities are available, derived from the same T1-w MRI scan (VBM, SBM, and quasi-raw). We focus this study on gray matter volumes (VBM). The model evaluation on OpenBHB is done on two private test sets (internal and external). The internal test set contains the same sites as training, the external contains unseen ones. 

\section{Experiments and Results}
\label{sec:experiments}

As network architecture, we employ the 3D implementations of ResNet-18 (33.2M parameters), AlexNet (2.5M parameters), and DenseNet-121 (11.3M parameters).
For comparison with~\cite{dufumier2022openbhb}, we use the Adam optimizer, with an initial learning rate of $10^{-4}$ decayed by a factor of $0.9$ every 10 epochs, and a weight decay of $5*10^{-5}$. We use a batch size of 32, and train for a total of 300 epochs.
Our trainings are implemented in PyTorch, and run on the Jean Zay supercomputer (NVIDIA V100 GPUs)\footnote{This work was granted access to the HPC resources of IDRIS under the allocation 2022-AD011013473 made by GENCI.} and on an NVIDIA A40 GPU, with a single training taking \url{~}24h.
For every model, we evaluate the mean absolute error (MAE) on both the internal and external test sets. Furthermore, the OpenBHB challenge computes a balanced accuracy (BAcc) for site prediction, training a logistic regression on the model representations. %
The final challenge score is then computed as $\mathcal{L}_c = \text{BAcc}^{0.3} \cdot \text{MAE}_{\text{ext}}$.\\

\noindent\subsection{Kernel function ablation study}
\begin{table}
    \centering
    \begin{tabular}{l l | l l l}
        \toprule
        Kernel & $\sigma$ / $\gamma$ & $\mathcal{L}^{y-aware}$ & $\mathcal{L}^{threshold}$ & $\mathcal{L}^{exp}$ \\ 
        \midrule
        Cauchy & 1 & 2.15 & 2.28 & 1.82 \\
               & 2 & 2.48 & 3.03 & 1.83 \\
        RBF & 1 & 2.43 & 2.63 & 1.58 \\
            & \textbf{2} & \textbf{1.82} & \textbf{1.74} & \textbf{1.54} \\
        \bottomrule
    \end{tabular}
    \caption{Ablation study of kernel functions, in terms of challenge's score.}
    \label{tab:kernel-ablation}
\end{table}
We test two different kernels: a Gaussian kernel $K_g(u) = \exp\left({-{||u||^2}/{2\sigma^2}}\right)$
and Cauchy kernel $K_c(u) = 1/(\gamma||u||^2 + 1)$. We perform an ablation study for the two different kernels and hyperparameters, employing a ResNet-18 model. Fig.~\ref{fig:kernel-ablation} shows the result. For each kernel, we report the metrics on the test set along with the final challenge score, for the three loss functions. Focusing on the final score, it's easy to see that a Gaussian kernel with $\sigma=2$ produces the best results for all losses (for readability, the final score is also reported in Tab.~\ref{tab:kernel-ablation}). 
This can be attributed to the overall lower error on the external set (Ext. MAE), showing that, with this setting, the models can generalize better. Furthermore, we also notice an overall lower balanced accuracy for site prediction, showing that this configuration is somewhat more robust to site noise.\\

\noindent\subsection{Comparison of contrastive regression losses}
In Tab.~\ref{tab:losses-comparison} we compare the results obtained with the different losses. Focusing on the aggregate score, the best results are obtained with $\mathcal{L}^{exp}$ (1.54). Furthermore, $\mathcal{L}^{exp}$ also outperforms the other losses in every evaluated metric. Most significantly, it shows the best generalization capability in the external test set, which, undoubtedly, is the most relevant result from a practical clinical perspective. 
On the internal test, we score a MAE of 2.55, which is also slightly better than the related literature on a similarly sized dataset with UKB~\cite{peng2021}. 
Interestingly, $\mathcal{L}^{exp}$ also shows the best robustness to site-related noise (with a BAcc of 5.1), which indicates that the learned space preserves the neuroanatomical features very well while also removing site noise. \\

\begin{table}
    \centering
    \begin{tabular}{r l l l c}
        \toprule
        Method & Int. MAE & BAcc & Ext. MAE & $\mathcal{L}_c$\\
        \midrule
        $\mathcal{L}^{y-aware}$ & \underline{2.66}\std{0.00} & 6.60\std{0.17} & 4.10\std{0.01} & 1.82 \\
        $\mathcal{L}^{thr}$ & 2.95\std{0.01} & \underline{5.73}\std{0.15} & 4.10\std{0.01} & 1.74 \\
        $\mathcal{L}^{exp}$ & \textbf{2.55}\std{0.00} & \textbf{5.1}\std{0.1} & \textbf{3.76}\std{0.01} & \textbf{1.54} \\
        \bottomrule
    \end{tabular}
    \caption{Comparison of contrastive losses.}
    \label{tab:losses-comparison}
\end{table}

\noindent\subsection{Final results on the OpenBHB challenge}
Finally, we report the ranking of $\mathcal{L}^{exp}$ of the OpenBHB leaderboard, testing also AlexNet and DenseNet-121. Tab.~\ref{tab:openbhb-results} shows the results. We compare with baseline models~\cite{dufumier2022openbhb} trained with the L1 loss, and with ComBat~\cite{fortin2018harmonization}, a site harmonization algorithm developed for MRIs. 
Our proposed $\mathcal{L}^{exp}$ achieves state-of-the-art performance on the final leaderboard, scoring the best final score and metrics on both the internal and external test set, with ResNet-18. The improvement observed in the external test is also reflected for both AlexNet and DenseNet compared to all baselines. For these models, the internal MAE reached by the L1 baseline is slighly lower than $\mathcal{L}^{exp}$. 
However, when looking at the other metrics, it is easy to see that this is due to more overfitting on the internal sites for the baseline. Lastly, regarding the balanced accuracy, we observe a significant improvement with respect to the L1 baseline, showing that $\mathcal{L}^{exp}$ possesses some debiasing capability towards site noise. Besides AlexNet, however, ComBat still achieves a lower accuracy, showing that there is room for improvement.

\begin{table}
    \centering
    \resizebox{\linewidth}{!}{%
    \begin{tabular}{r l l l l c}
        \toprule
        Method & Model & Int. MAE & BAcc & Ext. MAE & $\mathcal{L}_c$\\
        \midrule
        \multirow{3}{*}{Baseline ($\ell_1$)}
            & DenseNet & 2.55\std{0.01} & 8.0\std{0.9} & 7.13\std{0.05} & 3.34 \\
            & ResNet-18  & 2.67\std{0.05} & 6.7\std{0.1} & 4.18\std{0.01} & 1.86 \\
            & AlexNet & 2.72\std{0.01} & 8.3\std{0.2} & 4.66\std{0.05} & 2.21 \\
        \midrule
        \multirow{3}{*}{ComBat}
            & DenseNet & 5.92\std{0.01} & 2.23\std{0.06} & 10.48\std{0.17} & 3.38 \\
            & ResNet-18 & 4.15\std{0.01} & \textbf{4.5}\std{0.0} & 4.76\std{0.03} & 1.88 \\
            & AlexNet & 3.37\std{0.01} & 6.8\std{0.3} & 5.23\std{0.12} & 2.33 \\
        \midrule
        \multirow{3}{*}{$\mathcal{L}^{exp}$}
            & DenseNet & 2.85\std{0.00} & 5.34\std{0.06} & 4.43\std{0.00} & 1.84 \\
            & ResNet-18 & \textbf{2.55}\std{0.00} & {5.1}\std{0.1} & \textbf{3.76}\std{0.01} & \textbf{1.54} \\
            & AlexNet & 2.77\std{0.01} & 5.8\std{0.1} & 4.01\std{0.01} & 1.71\\
        \bottomrule
    \end{tabular}%
    }
    \caption{Final scores on the OpenBHB leaderboard. Reference results from~\cite{dufumier2022openbhb}.}
    \label{tab:openbhb-results}
\end{table}

\section{Conclusions and future works}
\label{sec:conclusions}
To the best of our knowledge, this is the first work proposing a contrastive learning loss for regression in medical imaging. We employ it to predict chronological 
brain age, using the OpenBHB multi-site MRI challenge.  We achieve state-of-the-art performance on the challenge score and in terms of generalization capabilities to unseen sites. This represents a valuable result for the scientific community, as building a robust brain age prediction model can help in better understanding neurodevelopmental or neurodegenerative disorders. We empirically demonstrate that contrastive learning is also capable of (partially) debiasing the site effects. In future, we will extend our experiments to include debiasing techniques, such as regularization terms \cite{barbano2023unbiased} 
 adapted for regression, to further improve the results.

\noindent \textbf{Compliance with ethical standards}
\label{sec:ethics}
This study was conducted using
human subject data, available in open access. Ethical approval was not required.

\bibliographystyle{IEEEbib}
\bibliography{strings,refs}

\begin{thebibliography}{10}

\bibitem{peng2021}
Han Peng et~al.,
\newblock ``Accurate brain age prediction with lightweight deep neural
  networks,''
\newblock {\em MedIA}, 2021.

\bibitem{wachinger2021detect}
Christian Wachinger et~al.,
\newblock ``Detect and correct bias in multi-site neuroimaging datasets,''
\newblock {\em MedIA}, 2021.

\bibitem{glocker2019machine}
Ben Glocker et~al.,
\newblock ``Machine learning with multi-site imaging data: An empirical study
  on the impact of scanner effects,''
\newblock in {\em MedNeurIPS Workshop}, 2019.

\bibitem{dufumier2022openbhb}
Benoit Dufumier et~al.,
\newblock ``Openbhb: a large-scale multi-site brain mri data-set for age
  prediction and debiasing,''
\newblock {\em NeuroImage}, 2022.

\bibitem{cole2017predicting}
James~H Cole et~al.,
\newblock ``Predicting brain age with deep learning from raw imaging data
  results in a reliable and heritable biomarker,''
\newblock {\em NeuroImage}, 2017.

\bibitem{Jonsson2019}
B.~A. Jonsson et~al.,
\newblock ``Brain age prediction using deep learning uncovers associated
  sequence variants,''
\newblock {\em Nature Communications}, vol. 10, no. 1, Nov. 2019.

\bibitem{khosla2020supervised}
Prannay Khosla et~al.,
\newblock ``Supervised contrastive learning,''
\newblock in {\em NeurIPS}, 2020.

\bibitem{tsai2022conditional}
Yao-Hung~Hubert Tsai et~al.,
\newblock ``Conditional contrastive learning with kernel,''
\newblock {\em arXiv:2202.05458}, 2022.

\bibitem{dufumier2022rethinking}
Benoit Dufumier et~al.,
\newblock ``Rethinking positive sampling for contrastive learning with
  kernel,''
\newblock {\em arXiv:2206.01646}, 2022.

\bibitem{chen2020simCLR}
Ting Chen et~al.,
\newblock ``A simple framework for contrastive learning of visual
  representations,''
\newblock in {\em ICML}, 2020.

\bibitem{barbano2023unbiased}
Carlo~Alberto Barbano et~al.,
\newblock ``Unbiased supervised contrastive learning,''
\newblock {\em arXiv:2211.05568}, 2022.

\bibitem{Graf2021}
Florian Graf et~al.,
\newblock ``Dissecting supervised contrastive learning,''
\newblock in {\em ICML}, 2021.

\bibitem{dufumier2021contrastive}
Benoit Dufumier et~al.,
\newblock ``Contrastive learning with continuous proxy meta-data for 3d mri
  classification,''
\newblock in {\em MICCAI}. Springer, 2021.

\bibitem{wang_understanding_2020}
Tongzhou Wang and Phillip Isola,
\newblock ``Understanding {Contrastive} {Representation} {Learning} through
  {Alignment} and {Uniformity} on the {Hypersphere},''
\newblock {\em ICML}, 2020.

\bibitem{fortin2018harmonization}
Jean~Philippe Fortin et~al.,
\newblock ``Harmonization of cortical thickness measurements across scanners
  and sites,''
\newblock {\em NeuroImage}, 2018.

\end{thebibliography}

\end{document}